\documentclass[11pt]{article}
\usepackage{moriondstyle,epsfig}

\def\Journal#1#2#3#4{{#1} {\bf #2}, #3 (#4)}

\def\be{\begin{equation}}
\def\ee{\end{equation}}
\def\bea{\begin{eqnarray}}
\def\eea{\end{eqnarray}}

\begin{document}
\vspace*{4cm}
\title{SUSY AND DARK MATTER CONSTRAINTS FROM THE LHC}

\author{ M.J. WHITE }

\address{Department of Physics, Cavendish Laboratory, J J Thomson Avenue, Cambridge CB3 0HE, UK}

\maketitle\abstracts{The ability of the LHC to make statements about the dark matter problem is considered, with a specific focus on supersymmetry. After reviewing the current strategies for supersymmetry searches at the LHC (in both CMS and ATLAS), some key ATLAS studies are used to demonstrate how one could establish that SUSY exists before going on to measure the relic density of a neutralino WIMP candidate. Finally, the general prospects for success at the LHC are investigated by looking at different points in the MSSM parameter space.}

\section{Introduction}
Several observations in the recent past have hinted at the existence
of dark matter, starting with the observation of anomalies in galaxy
rotational speeds, and culminating most recently in precision
measurements of the power fluctuations in the cosmic microwave
background (CMB). Indeed, the latter line of enquiry has proved
fruitful in estimating the amount of dark matter present in the
universe,  and the WMAP experiment~\cite{WMAP} has recently extracted
values for both the matter density and the baryon density of the
universe. Assuming the difference is due to the presence of dark matter, one obtains the following 2$\sigma$ range for the dark matter density:
\begin{equation}
0.093<\Omega_mh^2<0.129
\label{eq:WMAP}
\end{equation}
Furthermore, in order to be consistent with the observed structure of the universe, WMAP favours cold dark matter, or matter comprised of particles that are non-relativistic when galaxy formation starts. This naturally leads one to conclude that some sort of weakly interacting massive particle, or WIMP, is providing this dark contribution to our universe, and it is here that exciting astrophysics suddenly evolves into challenging and captivating particle physics!

The Large Hadron Collider (LHC) at CERN in Geneva will start taking
data next year and the purpose of this talk is to look at how much the
LHC will be able to say about the dark matter problem, from the direct
production and observation of WIMP candidates. Given the impossibility
of representing the state of the art in every potential scenario, the
focus throughout will be on the reasonably well motivated theory of
supersymmetry (SUSY). Even within supersymmetry, there are several
possibilities for WIMP candidates, and a decision to focus on the lightest neutralino is made here, as is the assumption that supersymmetry exists in nature under the guise of the Minimal Supersymmetric Standard Model (MSSM). On the astrophysical side, specific attention is given to the measurement of WIMP relic density, as opposed to any other property of the dark matter. 

In order to give as complete a picture of dark matter searches at the LHC as possible, I have chosen to address three important questions:
\begin{enumerate}
\item{How does one reconstruct SUSY models at the LHC?}
\item{How does one obtain the dark matter relic density from these models?}
\item{What are the general prospects for the LHC in different regions of parameter space?}
\end{enumerate}
A comprehensive review of earlier work is given by Battaglia et al~\cite{Battaglia}, and the focus of this talk is on the major developments that have occurred during the last year.
\section{Finding SUSY at the LHC}
\subsection{Supersymmetry}
In SUSY theories, all existing particles of the standard model have partners with opposite spin statistics called sparticles. Furthermore, one can impose a symmetry called R-parity under which the standard model particles are even whilst the SUSY particles are odd. This has two important phenomenological consequences:
\begin{enumerate}
\item{We will pair produce sparticles at the LHC.}
\item{The lightest sparticle (LSP) is absolutely stable.}
\end{enumerate}
Thus, the LSP is a natural WIMP candidate, and any consideration of
SUSY models is highly relevant to the search for dark matter. It is
noted that in different regions of the SUSY parameter space, one
obtains different LSP's, with possible options including the gluino,
sneutrino, gravitino and the lightest neutralino. The last particle in
this list is an admixture of the superpartners of the neutral SM gauge
bosons, and remains the subject of the majority of studies. Thus, it is the only SUSY candidate to be considered from now on.
\subsection{Reconstructing SUSY Models at the LHC}
The problem of finding SUSY at the LHC has occupied many researchers
in recent years, and only the most recent results will be presented
here. Any theory with as large a parameter space as the MSSM presents
a considerable challenge to experimenters, and it is therefore useful to
consider the problem in a stepwise fashion. 
\subsubsection{Step 1: Inclusive Searches}
\begin{figure}
\vskip 2.5cm
\centerline{
\psfig{figure=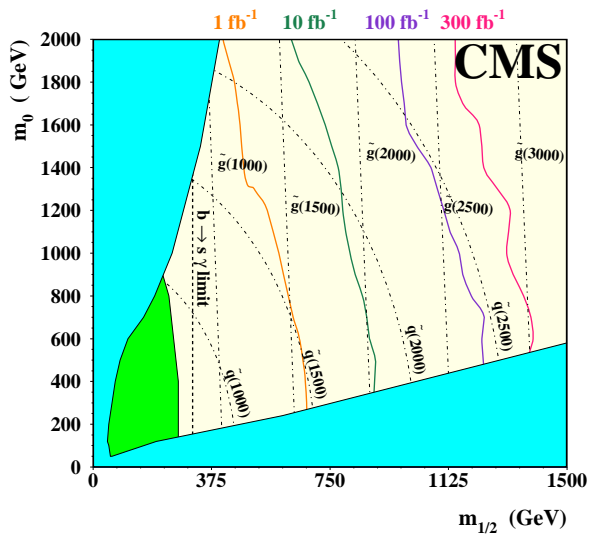,width=3.2in}
}
\caption{Search reach of the \em Jets \rm + $E_T^{miss}$ channel to
  obtain a 5 $\sigma$ discovery, as calculated in the framework of the
  CMSSM for the CMS detector, shown for different integrated
  luminosities. The dashed lines show fixed values of
  the squark and gluino masses.
\label{fig:CMS}}
\end{figure}

The first course of action is to perform inclusive searches that
exploit generic features of SUSY events. For example, the presence
of two invisible LSP's in each event leads to large amounts of missing
transverse energy, whilst squark and gluino decay ensure that jet
multiplicity is high. Furthermore, many SUSY processes produce
isolated leptons in conjunction with these other signatures. Hence,
inclusive searches in relevant channels are, naively, an excellent way to
discover R-parity conserving SUSY, even in the early days of data
taking. The problem is that the same could be said for other models
such as UED, and hence one needs to measure some more details of the
underlying model before a confident declaration of SUSY discovery can
be made believable. Nevertheless, inclusive searches will provide the
first key evidence that we have produced a WIMP candidate, and we see
in figure~\ref{fig:CMS} the search reach for the CMS detector, as
calculated in the CMSSM. One can infer that the prospects for
discovery of TeV-scale supersymmetry at the LHC are very good.
\subsubsection{Step 2: SUSY Parameter Extraction}
The second step in our search for SUSY involves the attempt to measure
some of the weak scale SUSY parameters. In the past, it has been assumed that the isolation of specific decay processes (a process known as `exclusive'
measurement) will be used to perform this step, although more
recent work has looked at combining exclusive and inclusive data to
improve the results; both will be reviewed here.
\begin{figure}
\vskip 2.5cm
\centerline{
\psfig{figure=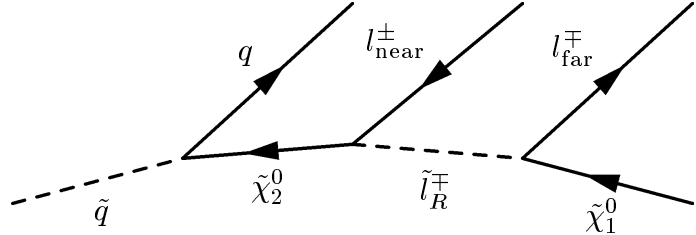,width=3.2in}
}
\caption{A popular cascade decay chain for use in exclusive analysis.
\label{fig:decay}}
\end{figure}

Consider firstly the problem of measuring sparticle masses. This is
non-trivial at the LHC for two reasons- all decay chains eventually
produce LSP's that leave the detector unseen and hence we do not see
all of the decay products, and we also do not know the centre of mass
frame at a hadron collider. However, one is always free to use clever
tricks. Consider the sparticle decay chain shown in figure~\ref{fig:decay},
which represents one side of an event in, for example, the ATLAS
detector (remember that sparticles are produced in pairs). If one is
in the rest frame of the squark, it is clear that the two decay
products- in this case a $\tilde{\chi}_2^0$ and a quark- cannot have an
invariant mass that exceeds the rest mass of the squark. The same
applies to the decay products of the $\tilde{\chi}_2^0$ in its own
rest frame. Ultimately, one will observe a jet and two leptons as the
visible products of the decay, and each possible invariant mass that
can be formed from these decay products has a theoretical maximum
which is given by a function of the four sparticle masses in the
chain. Reconstructing the masses is therefore simply a case of plotting
invariant mass distributions and looking for kinematic endpoints. The
chain shown in figure~\ref{fig:decay} has the advantage that the
standard model background is particularly small once one applies cuts
to select events with large missing energy and opposite sign same
flavour leptons. 
\begin{figure}
\vskip 2.5cm
\centerline{
\psfig{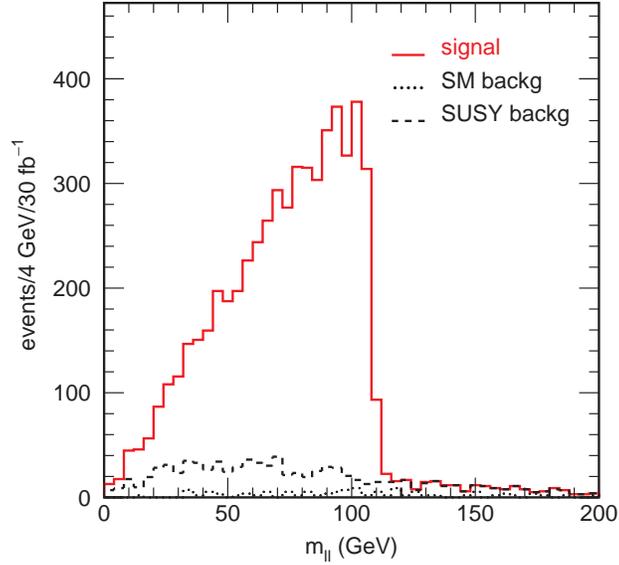}
}
\caption{ Example of a dilepton invariant mass distribution, showing the standard model and SUSY
  backgrounds. The signal is generated at a point in parameter space
  where the decay shown in figure~\ref{fig:decay} is open.
\label{fig:endpoint}}
\end{figure}

An example of a kinematic endpoint taken from the ATLAS Physics TDR~\cite{TDR} is shown in
figure~\ref{fig:endpoint}, and it is noted that further endpoints can
be seen in invariant mass distributions featuring combinations of the jet
and the leptons. Since each of the edge positions is a function of only
four masses, enough distributions can be obtained to solve for the
masses, and these can then be used to fit for the GUT scale SUSY
parameters if one assumes that the SUSY breaking scenario is known. 

Many studies have used this technique, though few of them have
directly addressed the problems of the approach. Apart from the fact that
the decay chain in figure~\ref{fig:decay} may not be open, there is a problem
arising from the fact that the kinematic endpoint equations are
sensitive to \em mass differences \rm rather than absolute
masses. Furthermore, the decay chain dealt with here will not be
determined unambiguously by the selection cuts- one would observe the
same visible decay products if the chain had other neutralinos in, or
if the slepton was right-handed rather than left-handed for
example. When trying to fit for GUT scale parameters, it is unclear whether a breaking scenario such as mSUGRA will prove sufficiently detailed- one must really try and fit in the general parameter space of the MSSM.

\begin{figure}
\vskip 2.5cm
\centerline{
\psfig{figure=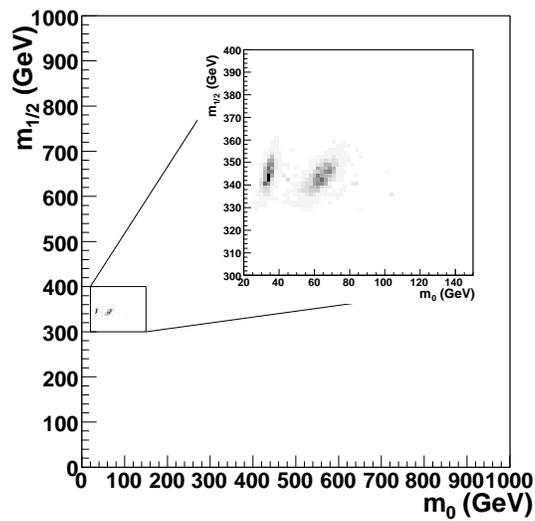,width=3.2in}
}
\caption{A sampling of the mSUGRA parameter space using endpoint data
  in conjunction with a measurement of the cross-section of events
  passing a missing $p_T$ cut of 500 GeV, with the effects of decay
  chain ambiguity included. The plot represents the posterior probability distribution of the mSUGRA parameters based on the assumed experimental input. Two regions result from the scan,
  reflecting a lack of knowledge about which slepton is involved in
  the decay chain.
\label{fig:ambig}}
\end{figure}

All of these problems are addressed in recent work performed by the
Cambridge group~\cite{me}, which combines inclusive and exclusive data
in a Markov Chain Monte Carlo sampling of SUSY parameter space. The
technique can easily be generalised to explore higher dimensional
parameter spaces than mSUGRA, and can include the effects of
ambiguities in decay chains (see figure~\ref{fig:ambig}). In addition,
mass measurements are improved by the fact that the inclusive
information is sensitive to the mass scale. 
The process described by the Cambridge group should be enough to determine
properties of the SUSY Lagrangian, although it is noted that one ought
to measure the spins of particles in order to make sure that we have
observed SUSY rather than, for example, UED. This has been considered in more
detail by Barr~\cite{Barr}.
\subsection{Determining the Dark Matter Relic Density From SUSY
  Measurements}
In general, one finds that too many neutralinos are produced after the
big bang, and we therefore require some kind of annihilation mechanism
to bring the density down to within the limits set by astrophysical
observation. There are four main mechanisms that can occur (at least within the mSUGRA framework):
\begin{enumerate}
\item Slepton exchange. This is suppressed unless the slepton masses
  are lighter than approximately 200 GeV. 
\item Annihilation to vector bosons. This can occur if the neutralino
  LSP acquires a significant wino or higgsino component.
\item Co-annihilation with light sleptons. This occurs when there are
  suitable mass degeneracies in the sparticle spectrum.
\item Annihilation to third-generation fermions. This is enhanced when
  the heavy Higgs boson $A$ is almost twice as massive as the LSP.  
\end{enumerate}
\begin{figure}
\vskip 2.5cm
\centerline{
\psfig{figure=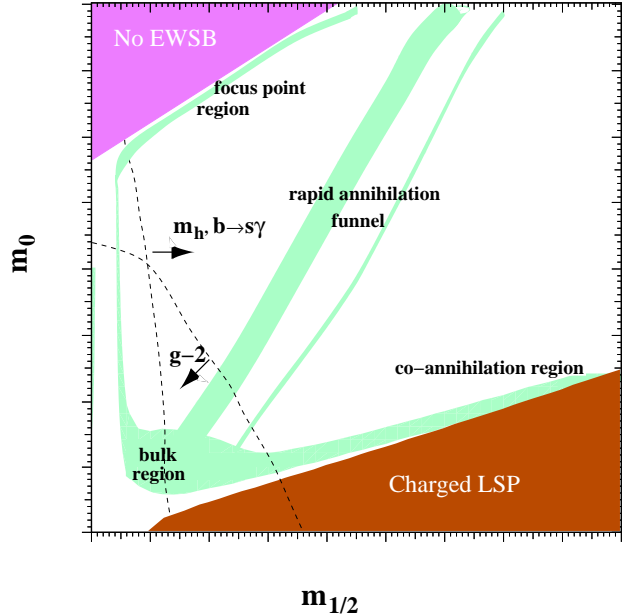,width=3.2in}
}
\caption{A schematic plot of the mSUGRA $m_0-m_{1/2}$ plane showing
  the regions that are consistent with the WMAP relic density
  constraint, taken from Nojiri et al. The bulk region features
  annihilation through slepton exchange, the focus point region
  involves an enhanced annihilation to vector bosons, the funnel
  region involves enhanced annihilation to third-generation fermions,
  and the co-annihilation region is that in which mass degeneracies
  occur in the sparticle spectrum. 
\label{fig:regions}}
\end{figure}
 
Although these mechanisms can, and do, occur simultaneously in different regions of parameter space, imposing the WMAP constraint tends to give
allowed regions in which one of these mechanisms is
dominant (see figure~\ref{fig:regions}). It is noted that more recent work examining the current state of the mSUGRA parameter space using astrophysical constraints along with other information has been performed by Allanach and Lester~\cite{Allanach} and, later still, Trotta et al~\cite{Trotta}, and that the effects of imposing the WMAP constraint on models more general than mSUGRA have been described by Belanger et al~\cite{Belanger}. We see that LHC experimenters will need to measure enough information to determine which region Nature has chosen and thus, naively, one must be able to determine the LSP mass, the masses of other light sparticles, the mass of the heavy Higgs boson $m_A$, and the components of
the neutralino mixing matrix:
\begin{eqnarray}
{\cal M}=\left(\begin{array}{cccc}
  M_1       &      0          &  -m_Z \cos\beta s_W  & m_Z \sin\beta s_W \\[2mm]
   0        &     M_2         &   m_Z \cos\beta c_W  & -m_Z \sin\beta c_W\\[2mm]
-m_Z \cos\beta s_W & m_Z \cos\beta c_W &       0       &     -\mu        \\[2mm]
 m_Z \sin\beta s_W &-m_Z \sin\beta c_W &     -\mu      &       0
                  \end{array}\right)\
\label{eq:massmatrix}
\end{eqnarray}
where $M_1$ and $M_2$ are the U(1) and SU(2) gaugino masses, $\mu$ is
the Higgsino mass parameter, tan$\beta$ is the ratio of the vacuum
expectation values of the two Higgs doublets and the other parameters
are all from the standard model. Most significantly, we see that a
complete knowledge of neutralino mixing requires some knowledge of the
SUSY Higgs sector. 

There are essentially two strategies for determining this long list of
information; try and fit a GUT scale SUSY model or (more
realistically) aggressively target the weak scale parameters relevant
to the relic density calculation. An excellent example of the second
approach is that recently published by Nojiri, Polesello and
Tovey~\cite{Tovey}. They use an existing study of an mSUGRA benchmark
point in the co-annihilation region (where the third mechanism in the
previous list is the most significant), but perform an analysis within
the framework of a general MSSM. 

Their starting point is the exclusive analysis presented above- they
use endpoint data to constrain sparticle masses (though they do not
consider the problems tackled by Lester, Parker and White~\cite{me}). They then use these mass
values to constrain the neutralino mixing matrix, though they only
obtain three (of four) neutralino masses and hence lack one parameter
to constrain the matrix. They thus obtain only tan$\beta$ dependent
values of the mixing parameters but, nevertheless, manage to establish
that the LSP is predominantly bino. Having established from the mass
spectrum information that co-annihilations are likely to be important,
they next set about trying to constrain the slepton sector using a
ratio of branching fractions that is sensitive to the stau mixing
parameters: $BR(\tilde{\chi}_2^0 \to \tilde{l}_Rl)/BR(\tilde{\chi}_2^0
\to \tilde{\tau}_1\tau)$. Again, their results are tan$\beta$
dependent. 

Finally, they consider constraints on the Higgs sector, although these are challenging due to the fact that their benchmark point is in a
region in which ATLAS is not expected to observe anything other than
the lightest (SM-like) Higgs boson. They obtain a relic density
distribution as a function of $m_A$, shown in figure~\ref{fig:relic},
but can improve their measurement by placing a lower limit of 300 GeV
on $m_A$ due to its non-observation in cascade decays. This
well-motivated assumption gives them a massive improvement in their
control over the relic density, and they obtain a final value of:
\begin{figure}
\vskip 2.5cm
\centerline{
\psfig{figure=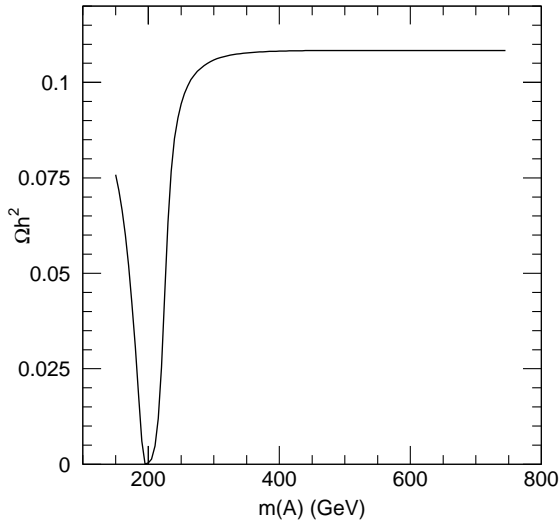,width=3.2in}
}
\caption{The relic density of neutralino dark matter as a function of
  the pseudoscalar Higgs mass, as obtained by Nojiri et al.
\label{fig:relic}}
\end{figure}
\begin{equation}
\Omega_\chi h^2 = 0.108 \pm 0.01 (stat + sys)^{+0.00}_{-0.002}
(M(A)) ^{+0.001}_{-0.011} (\tan\beta) ^{+0.002}_{-0.005} (m(\tilde{\tau}_2))
\end{equation}
\subsection{General Prospects for Dark Matter Observation at the LHC}
Having considered in detail one example of a dark matter search at the
LHC, it is worth exploring whether the success encountered there is likely
to be repeated in other regions of the parameter space, or whether it
was specific to the chosen benchmark point. It is hard to be totally
general here, but there are nevertheless some generic remarks that it
is possible to make.

Firstly, we will always need to measure the masses of the lightest
sparticles and the mixing parameters for the lightest neutralino. This
ultimately leads one to conclude that the LHC will perform best in
regions where light sparticles are copiously produced in cascade
decays.

Secondly, it is noted that if co-annihilations are important, the mass
differences in cascade decays will be small, and hence the visible
(SM) decay products produced in such decays will have low transverse
momentum and may be missed by the CMS and ATLAS detectors (which
typically will only function well down to a $p_T$ of approximately 5
GeV). Thus the very mechanism that allows SUSY to produce a consistent
picture of dark matter may scupper our chances of measuring it! In
such a case, one would hope to be able to constrain the SUSY
Lagrangian from other measurements, but the LHC may prove insufficient to accomplish this.

Finally, we have seen that tan$\beta$ is an important quantity to know
if we want to calculate the dark matter relic density, and this will
always be difficult to measure at the LHC.

To put these points on a firmer footing, it is useful to refer to a recent study by Baltz et al~\cite{pros} (reprising some themes considered earlier by Allanach et al \cite{Allanach2}) that looks at points in different regions of the mSUGRA
parameter space (though they are analysed in the framework of the MSSM), each of which has a dominant LSP annihilation mechanism
that is one of the four introduced earlier. Their first point is
similar to that studied by Nojiri et al, and they reach similar
conclusions. It is therefore more interesting to consider their points
LCC2 and LCC4.

\begin{figure}
\vskip 2.5cm
\centerline{
\psfig{figure=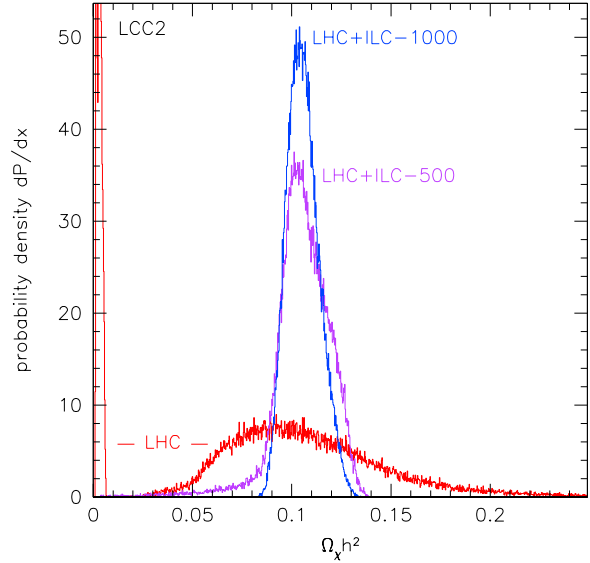,width=3.2in}
}
\caption{The posterior probability distribution of the dark matter relic density
  as determined from collider observables at the LHC and/or linear
  collider, calculated for a point in the focus point region.
\label{fig:lcc2}}
\end{figure}

LCC2 is in the `focus point region' (where LSP annihilation occurs via
enhanced annihilation to vector bosons). Their point has a large value
of $m_0$, and the resulting squarks and sleptons are too heavy to be
observed at the LHC. The result is that one does not obtain enough
information to constrain the neutralino mixing matrix, and hence
cannot constrain the relic density, shown clearly in figure~\ref{fig:lcc2}
which plots the posterior probability distribution for
$\Omega_{\chi}h^2$ obtained using a Markov Chain Monte Carlo sampling
method. The same figure shows the improvement that results if one has
access to the linear collider (running at both 500 and 1000 GeV).

A similar lack of constraint is observed in the funnel region, though
for different reasons. Here, annihilation proceeds via a Higgs
resonance (due to $m_A$ being roughly twice the LSP mass), and a
knowledge of the $A$ decay width is needed to constrain the
relic density. This cannot be done at the LHC, though it can be
measured at a linear collider. It is worth noting that the study also
examines the interplay between astrophysical direct/indirect search
experiments and collider results, and it is hard to escape the
conclusion that both sets of data will prove essential if we are to
develop a comprehensive understanding of the dark matter problem.
\begin{figure}
\vskip 2.5cm
\centerline{
\psfig{figure=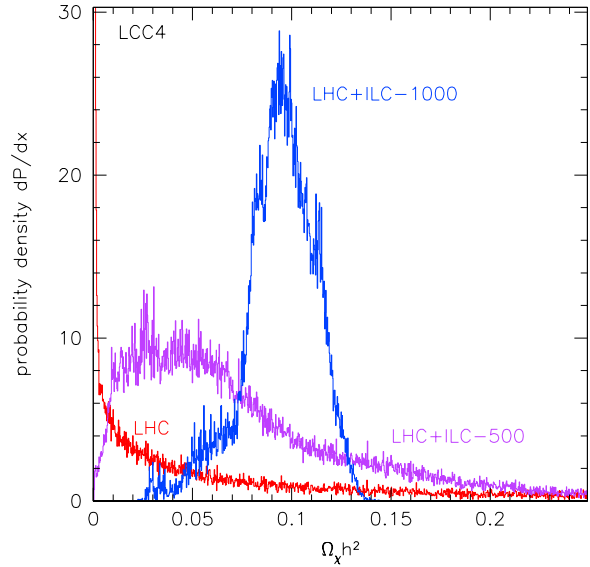,width=3.2in}
}
\caption{The posterior probability distribution of the dark matter relic density
  as determined from collider observables at the LHC and/or linear
  collider, calculated for a point in the funnel region.
\label{fig:lcc4}}
\end{figure}

\section{Summary}
Physicists are currently entering one of the most exciting periods in
the history of the subject. Not only do we believe that there is dark
matter, but there is a wide array of experiments just around the
corner that have the potential to explain what is currently a fascinating mystery of cosmic proportions. 

I have briefly reviewed recent work that explains how to use the LHC to learn
about dark matter, using supersymmetry as an example. It has been seen
that the LHC is an excellent discovery machine, with a wide search
reach for observing SUSY WIMP candidates in inclusive
channels. Whilst it remains true that pinning down the precise nature of the SUSY model will prove harder, recent work allows one to make more model independent
statements in this area than were previously possible.

It has been shown that the LHC may be capable of determining the dark
matter relic density with a precision of approximately 10\% but that
this is highly dependent on the underlying SUSY model. There are
indeed very specific reasons why the LHC might fail, and it is
possible that the LHC will prove insufficient to completely constrain
WIMP properties.

Finally, it is worth noting that there are questions which a collider
can never address. Specifically, colliders alone will never determine
how much of the observed astrophysical dark matter is comprised of
WIMPS, nor reveal anything about the dark matter spatial and velocity
distributions. It is also worth remembering that we would only know we have produced a WIMP candidate if we know its lifetime. For these reasons, direct and indirect experiments are
entirely complementary to the collider programs at the LHC and/or
linear collider.

\section*{Acknowledgements}
I am hugely grateful to the ATLAS Speakers Committee for inviting me
to give this talk, and to Ms Patricia Chemali for her help in
securing funding. Many thanks must go to the European Union ``Marie
Curie'' programme for their financial support, and to PPARC for
allowing me to go to CERN prior to the conference to receive feedback.

\end{document}